\documentclass[useAMs,usenatbib,a4paper]{mn2e}
\usepackage{savesym}
\usepackage{graphicx}
\expandafter\let\csname equation*\endcsname\relax
  \expandafter\let\csname endequation*\endcsname\relax 
\usepackage{subfig}
\usepackage{amsmath}
\usepackage{amssymb}
\usepackage{verbatim}
\usepackage{array}
\usepackage{times}
\usepackage[total={17.8cm,24.0cm},centering]{geometry} 

\newcommand{\be}{\begin{equation}}
\newcommand{\beq}{\begin{equation}}
\newcommand{\ee}{\end{equation}}
\newcommand{\eeq}{\end{equation}}
\newcommand{\eea}{\end{eqnarray}}
\newcommand{\bea}{\begin{eqnarray}}

\newcommand{\dd}{\partial}

\newcommand{\m}{\mathrm}
\newcommand{\Pm}{{\rm Pm}}

\newcommand\bb[1] { \mbox{\boldmath{$#1$}} }

\title[An instability induced by a temperature sensitive $\alpha$ ]{An accretion disc instability induced by a temperature sensitive $\alpha$ parameter}
\author[William J. Potter and Steven A. Balbus]{William J. Potter\thanks{E-mail:
will.potter@astro.ox.ac.uk (WJP)} and Steven A. Balbus
\\
Oxford Astrophysics. Denys Wilkinson Building, Keble Road, Oxford, OX1 3RH, United Kingdom}
\begin{document}

\date{}

\pagerange{\pageref{firstpage}--\pageref{lastpage}} \pubyear{2012}

\maketitle

\label{firstpage}

\begin{abstract}
In the standard thin disc formalism the dimensionless $\alpha$ parameter is usually assumed to be constant.   However, there are good theoretical reasons for believing, as well as evidence from simulations, that $\alpha$ is dependent on intrinsic disc properties. In this paper we analyse the conditions for the stability of a thin accretion disc in which $\alpha$ is a function of the magnetic Prandtl number, the ratio of collisional viscosity to resistivity. In the inner disc, where the free electron opacity and radiation viscosity dominate, the disc is unstable if $\alpha$ is proportional to the magnetic Prandtl number with an exponent, $n$, and $6/13<n<10/3$.  This is within the range of values for the power-law index found in MHD simulations with simple energetics.  We calculate the evolution of the unstable disc within the $\alpha$ formalism and show that the physically accessible solutions form a limit cycle, analogous to the behaviour seen in recurrent dwarf novae. It is noteworthy that the time-dependent global behaviour of the instability results in cyclic heating of the inner section of the disc, when parameters appropriate for an X-ray binary system are used.  We calculate a model spectrum of the disc in the flaring and quiescent states and show that the behaviour is compatible with X-ray observations of the thermal accretion disc in flaring X-ray binary systems.

\end{abstract}

\begin{keywords}
accretion, accretion discs -- instabilities -- (stars:) novae, cataclysmic variables -- (stars:) binaries: general
\end{keywords}

\section{Introduction}

Accretion discs are noteworthy for their remarkable variability.  Cataclysmic variables, in particular dwarf novae (DN), are among the most intensively studied variable disc systems (e.g. \citealt{2002apa..book.....F}). These are generally analysed using the so-called $\alpha$-formalism, in which the turbulent stress is assumed to be linearly proportional to the pressure (on dimensional grounds), the proportionality constant being denoted as $\alpha$.  Within the $\alpha$-formalism (and somewhat more generally, e.g. \citealt{1999ApJ...521..650B}), the combination of mass and angular momentum conservation leads to a diffusion equation, the characteristic timescale of which is often referred to as the \lq{}viscous time\rq{}.  The viscous time is a measure of the time for a fluid element to drift radially inward across a characteristic disc radius, and is generally the longest time scale associated with the accretion process.  It is then convenient to assume that the more rapid heating and cooling processes always maintain an effective thermal equilibrium while the disc itself slowly evolves. 

In $\alpha$ disc theory, an infinitesimal surface density perturbation satisfies its own linearised diffusion equation. Under conditions in which the diffusion coefficent for this equation is negative, this leads to dynamical instability. Applied to classical DN theory, negative diffusion occurs only at temperatures at which hydrogen is thermally ionized (and the form of the radiative opacity correspondingly complex). The ensuing {\em nonlinear} behaviour corresponds to a limit cycle about the unstable equilibrium, which is identified with the repeated outbursts characteristic of DN.  

CV systems are not the only binary accretors to show variability.  Equally well-known is a class of objects known as X-ray transients \citep{2006ARA&A..44...49R}.  The evolution of these sources is quite complex, and associated with higher temperatures {\em not} characteristic of hydrogen ionisation. Indeed, in the regime of interest, the opacity is often modelled by a simple Kramers' law in the outer part of the disc and by electron scattering in the inner regions \citep{2002apa..book.....F}. This suggests another instability mechanism could be at work.

The observed changes in the radiative states of X-ray transients strongly suggest that the inner region of a \cite{1973A&A....24..337S} $\alpha$ disc is disrupted when a transition from the thin disc state occurs.  The difficulty is that there seems to be no obvious reason why a viable, apparently stable, classical $\alpha$ disc should suddenly and spontaneously break down. Two suggestions that have been studied are i.) that evaporation by a (disc-fed) tenuous corona may preferentially remove the inner disc \citep{2000A&A...361..175M}; 
ii.) that the existence of another type of accretion solution, hot and diffuse, somehow induces the disc to make the transition within some characteristic radius \citep{1995ApJ...452..710N}.    
In this paper we revisit the problem of X-ray transients within the framework of $\alpha$ disc theory,  incorporating a particularly important  finding of MHD turbulence theory, which offers a natural mechanism for inducing disc transitions. 

In classical disc theory, the $\alpha$ parameter is regarded as a simple mathematical constant. But the true behaviour is obviously more complex than this:  consider, for example, the magnetic response of a cool, resistive gas versus a hot, fully-ionized plasma.  The idea we shall explore in this paper is that the ratio of the (microscopic) viscosity to resistivity, known as the magnetic Prandtl number, $\Pm$, could influence the saturated value of the stress tensor for MHD turbulence \citep{1998RvMP...70....1B}.  There is strong evidence for this, at least in the regimes accessible to numerical simulation (\citealt{2007MNRAS.378.1471L}, \citealt{2007A&A...476.1123F}, \citealt{2009ApJ...707..833S}, \citealt{2010A&A...516A..51L} and \citealt{2011ApJ...730...94S}). When $\Pm$ is near unity, the radial-azimuthal component of the stress increases monotonically with this number.  Turbulent dynamos, in particular, behave very differently depending upon whether $\Pm$ is greater or less than one \citep{2004ApJ...612..276S}. Large $\Pm$ dynamos are easily excited in driven turbulent simulations. Small $\Pm$ dynamos are much more difficult to excite, show lower rms levels of fluctuations, and much less spatial coherence. It would be surprising if MHD turbulent discs were completely indifferent to all this, yet this aspect of disc theory is in fact widely ignored.

In liquid metals and stellar interiors, $\Pm \ll 1$.  In X-ray discs, however, it is likely that regions with $\Pm \ll 1$ and $\Pm \gg 1$ are both present \citep{2008ApJ...674..408B}.  The $\Pm = 1$ transition radius typically occurs at 50--100 Schwarzschild radii.  CV systems, associated with white dwarfs, have no such Pm-transition radius.  We are led, therefore, to consider the behaviour of accretion discs for which the normalised stress (i.e. $\alpha$) is not constant, but a dynamic variable in its own right, whose value depends upon $\Pm$. 

The thermal, viscous and gravitational instability criteria for a thin disc, with a variable $\alpha$ depending on the magnetic Prandtl number, have recently been investigated by \cite{2011ApJ...727..106T}. They calculated the mathematical conditions for a density perturbation to become unstable in a gas or radiation pressure dominated disc, with either a dominant Kramers' or electron scattering opacity. In this paper we will focus on understanding the physical mechanism behind this viscous instability and investigate its time-dependent global behaviour, specifically, in relation to the observed flaring behaviour of X-ray binary systems.  We also highlight the role of radiation viscosity.

\section {Analysis}

\subsection{Thin disc equations}

We make use of the standard thin disc alpha formalism \citep{2002apa..book.....F}. These equations are based on the assumption that the disc is cool (the sound speed is much less than the rotation speed; in this sense thin), and in thermal and dynamical equilibrium.  The gas is thus on nearly circular orbits, and the rotational free energy is locally dissipated as heat and then efficiently radiated.   Another important assumption of this model is that the dissipation is localized close to the midplane so that the vertical flux of radiation is nearly constant with height above the dissipation layer; the dynamics could be very different if the dissipation were well above the midplane.  For ease of reference and to establish notation we review the fundamental equations.

\subsubsection{Coordinates and dynamics}

We use the standard $R$, $\phi$, $z$ cylindrical coordinates for the disc, where $R$ is the distance from the rotation axis, $\phi$ the azimuthal angle, and $z$ the vertical coordinate.  The gas moves in circular orbits about a central mass $M$. The Keplerian angular velocity $\Omega$ is a function of $R$ only and given by
\be
\Omega^2(R) =\frac{GM}{R^3},
\ee
where $G$ is the usual gravitational constant.  The velocity vector is $\bb{v}$ with components $v_R$, $v_\phi$, $v_z$.  Fluctuations from Keplerian flow will be denoted by $\delta v_R$, etc.  

The gas pressure and mass density are denoted as $P$ and $\rho$ respectively. They are related by the usual ideal gas equation of state
\be
P = \rho c_S^2 \equiv \frac{\rho k T}{m},
\ee
where $c_S$ is the isothermal sound speed and $m$ is the mean mass per particle. The magnetic field, which appears here only as part of the stress tensor, is denoted as $\bb{B}$.  The magnetic energy density is much less than the thermal energy density and an order of magnitude yet smaller than the rotational energy density. The Alfv\'en velocity is
\be
\bb{v_A}\equiv \frac{\bb{B}}{\sqrt{\mu_0\rho}},
\ee
where $\mu_0$ is the vacuum permittivity.  

The disc is presumed to be turbulent due to the magnetorotational instability (hereafter MRI, \citealt{1998RvMP...70....1B}), and the dominant $R\phi$ component of the turbulent stress tensor is $T_{R\phi}$. It is formally given by
\be
T_{R\phi} = \langle \rho(\delta v_R \delta v_\phi - v_{AR}v_{A\phi} ) \rangle,
\ee
where the angle brackets denote a suitable average (e.g., \citealt{1999ApJ...521..650B}). We shall henceforth refer to $T_{R\phi}$ as ``the stress''.   

The stress has dimensions of pressure, and it is natural to scale this dependence out, in the process defining a dimensionless quantity $\alpha$:
\be
T_{R\phi} \equiv \alpha P, \label{Trphi},
\ee
where $P$ is evaluated at the midplane.  It is also useful to introduce a form of the stress with dimensions of velocity, i.e., with the midplane density appropriately scaled out.  Dividing $T_{R\phi}$ by a locally averaged density $\langle\rho\rangle$, we define $W_{R\phi}$:
\be
W_{R\phi} \equiv \frac{T_{R\phi}}{\langle \rho \rangle} \equiv \alpha c_S^2.
\ee
The $\alpha$ parameter is, of course, {\em not} a constant in our analysis, but will depend upon the magnetic Prandtl number $\Pm$
(see below).
A strictly isothermal disc in hydrostatic equilibrium has a vertical Gaussian profile with a scale height $H=\sqrt{2}c_S/\Omega$. We will follow the long-established practice of multiplying by $2H$ to go from volume-specific to area-specific quantities in a thin disc.  

The combination of mass and angular momentum conservation leads to an evolutionary equation for the midplane-to-surface disc column density $\Sigma$ (\citealt{1981ARA&A..19..137P}; \citealt{1999ApJ...521..650B}):
\be
\frac{\dd\Sigma}{\dd t} = \frac{2}{\sqrt{GM}R} \frac{\dd{ }}{\dd R} \left[ R^{1/2} 
\frac{\dd{ }}{\dd R} (\Sigma R^2 W_{R\phi})\right], \label{Diff}
\ee
describing the time evolution of the system. It is apparent that if 
\be\label{unstable}
\frac{\dd(\Sigma W_{R\phi})}{\dd\Sigma} < 0,
\ee
the perturbations $\delta\Sigma$ of the column density will obey a diffusion equation with a {\em negative} diffusivity and are thus unstable.  The partial derivative with respect to $\Sigma$ is taken assuming radiative equilibrium, as discussed below.  

\subsubsection{Radiative considerations}

The volume specific rate at which energy is extracted from differential rotation is $-T_{R\phi}\partial \ln \Omega/ \partial \ln R$ \citep{1998RvMP...70....1B}. If this energy is then locally dissipated and radiated, as is generally assumed, the equation of thermal energy balance is
\be\label{eins}
- HT_{R\phi}{\m{d}\Omega\over \m{d}\ln R} =  \sigma T_s^4,
\ee
where $T_{s}$ is the surface temperature and $\sigma$ is the Stefan-Boltzmann constant (a factor of 2 has been cancelled on both sides). In turn, $T_{s}$ is related to the central midplane temperature $T_{c}$ by the equation \citep{2002apa..book.....F}
\be
T_c^4= \frac{3\tau}{4} T_s^4, \label{Tsurf}
\ee
where $\tau$ is the midplane-to-surface optical depth. 

The opacity is assumed to be a combination of electron scattering, bound-free and free-free absorption. If the disc is not fully ionised then bound-free absorption will be considerably larger than free-free absorption.   The absorption opacity is approximated here by its Kramers form. The total opacity, $\kappa$, is then \citep{2002apa..book.....F}  
\be
\kappa=(0.4+5\times 10^{24}\rho T_{c}^{-7/2}) \m{g\, cm}^{-2}.
\ee
The radius at which the electron scattering opacity and bound-free Kramers' opacity are equal is given by \citep{2002apa..book.....F}
\be
R_{\tau}\approx2.5\times10^{7}\left(\frac{\dot{m}}{10^{16}\m{gs}^{-1}}\right)^{2/3}\left(\frac{M}{M_{\odot}}\right)^{1/3}\m{cm}.
\ee
where $\dot{m}$ is the mass accretion rate and $M$ is the mass of the central object. Measuring the accretion rate in terms of the Eddington luminosity $L_{\m{Edd}} =1.26\times10^{38} M/M_\odot$, and distances in units of the Schwarzschild radius $r_{s}=2GM/c^2$ this becomes
\be
R_{\tau}\approx 480\left(\frac{\dot{m}c^{2}}{L_{\m{Edd}}}\right)^{2/3} r_{s}.
\label{relect}
\ee
Within this radius, the electron scattering opacity is dominant.   It is relevant to discs with accretion rates approaching the Eddington value,
$L_{\m{Edd}}/c^2$.    We show below that an electron scattering disc is more easily destabilised than a Kramers disc. 
The midplane-to-surface optical depth of the disc is defined by  
\be\label{zwei}
\tau = \Sigma \kappa.
\ee

With both radiative and dynamical equilibrium holding for the unperturbed disc solution, the instability criterion (\ref{unstable}) can be shown to be equivalent to \citep{2002apa..book.....F}
\be\label{unstable2}
\frac{\dd T_s}{\dd\Sigma}<0, 
\ee
a particularly convenient form to use for our purposes.






\subsubsection{Resistivity and viscosity}

We use the Spitzer (1962) resistivity in a form adopted by Balbus \& Henri (2008):
\be
\eta = \frac {5.55\times 10^{11} \, \ln \Lambda_{\m eH}} {T^{3/2}}\ {\m cm}^2\, {\m s}^{-1}
\ee 
where $\Lambda_{\m eH}$ is the electron-proton Coulomb collision factor.   Similarly, the Coulomb viscosity is
\be
\nu_C = \frac  {1.6\times10^{-15} T^{5/2}}{\rho \ln\Lambda_{\m HH}} \ {\m cm}^2\, {\m s}^{-1}
\ee
The corresponding radiation viscosity is \citep{1984oup..book.....M}:
\be
\nu_\m{ Rad} = 6.73\times 10^{-26}\, \frac {T^4}{\kappa\rho^2} \ {\m cm}^2\, {\m s}^{-1},
\ee
where $a$ is the standard radiation constant and $c$ the speed of light.
(There is an additional factor of $10/9$ for free electron scattering \citealt{1992ApJ...384..115L}.)
Their ratio is
\be
\frac{\nu_C}{\nu_\m {Rad}} = \left[\frac{T}{2.4\times10^6}\right]^{-3/2} \kappa\rho 
\label{nuc/nur}
\ee
where $\kappa\rho$ is in cm$^{-1}$.   (For numerical calculations, we take a nominal value of $\sqrt{40}\simeq6.3$
for any Coulomb logarithm.)  

At temperatures above several million Kelvin, well within our range of interest, radiation viscosity becomes important.  Thus, we require two 
Prandtl numbers, one that is radiative:
\be
{\m Pm}_{\, \m {Rad}} = 1.9\times 10^{-38} \, \frac{T^{11/2}} {\kappa\rho^2},
\label{Pmradvisc}
\ee
and one that is Coulombic:
\be
{\m Pm}_{\ C} = 7.2\times 10^{-29} \frac {T^4}{\rho}
\ee
In either form, Pm is very temperature sensitive, the radiative version exceptionally so: the dependence is 
$T^9/\rho^3$ for Kramers opacity.

\section{Instability criteria}




\subsection{Coulomb viscosity}

We next turn to calculating the conditions for the instability criterion (\ref{unstable2}) to occur within the standard $\alpha$ formalism. Substituting (\ref{Trphi}) into (\ref{eins}) gives us the following scaling for $T_{s}$ at a given radius, 
\be
T_{s}^{4} \propto \alpha T_{c} \Sigma.
\label{tc}
\ee
Taking the partial derivative of both sides with respect to $\Sigma$ at constant radius 
\be
4T_{s}^{3} \frac{\partial T_{s}}{\partial \Sigma} \propto \frac{\partial (\alpha T_{c} \Sigma)}{\partial \Sigma}.
\label{inst2}
\ee
Next, use (\ref{Tsurf}) in (\ref{tc}) to obtain 
\be
T_{c}\propto (\alpha \tau \Sigma)^{1/3}, \label{radequi}
\ee
and substitute this into (\ref{inst2})
\be
4T_{s}^{3} \frac{\partial T_{s}}{\partial \Sigma} \propto \frac{\partial (\alpha^{4/3} \Sigma^{4/3} \tau^{1/3})}{\partial \Sigma},
\ee
\be
T_{s}^{3} \frac{\partial T_{s}}{\partial \Sigma} \propto \alpha^{1/3} \Sigma^{1/3} \tau^{-2/3}\left(4\Sigma \tau\frac{\partial \alpha}{\partial \Sigma}+4\alpha \tau+ \alpha \Sigma \frac{\partial \tau}{\partial \Sigma}\right).
\ee
Only the bracketed quantity containing derivatives on the RHS can be negative. Our general instability criterion becomes
\be
\frac{\partial \ln \alpha}{\partial \ln \Sigma}+\frac{1}{4}\frac{\partial \ln \tau}{\partial \ln \Sigma} + 1<0.
\label{geninst}
\ee
This shows that if $\alpha$ is a sufficiently steeply changing function of the disc properties it can cause an accretion instability in the disc. This is analogous to the instability in DN accretion discs where the opacity is a steep function of $T_{c}$ around the ionisation temperature of hydrogen, which can lead to a transition between disc states \citep{1983MNRAS.205..359F}. The instability induced by a variable $\alpha$ may be relevant in explaining the observed transitions between spectral states of X-ray binaries. If the optical depth can be expressed in power law form, $\tau \propto \Sigma \rho T_{c}^{m}$, we can calculate the optical depth as a function of $\Sigma$ using (\ref{radequi}):
\be
\tau \propto \Sigma^{ \frac{11+2m}{7-2m} }\alpha^{\frac{2m-1}{7-2m}}.
\ee
The instability criterion in (\ref{geninst}) becomes
\be
\left(\frac{9-2m}{7-2m}\right) \frac{\partial \ln \alpha }{\partial \ln \Sigma}+\left(\frac{13-2m}{7-2m}\right) <0.
\label{dwarfinst}
\ee
In the case of constant $\alpha$ we recover the instability criterion for the opacity in dwarf novae ($7/2<m<13/2$). However, if we choose the electron scattering opacity ($\tau \propto \Sigma$), which is relevant for the inner region of the accretion disc (\ref{relect}) and allow for a variable $\alpha$, we can immediately calculate the instability condition from (\ref{geninst})
\be
\frac{\partial \ln \alpha }{\partial \ln \Sigma}<-\frac{5}{4}.
\label{instkram}
\ee
Thus, instability is present if $\alpha$ decreases sufficiently steeply with increasing surface density. Coincidentally, the same instability criterion is also valid in the case of Kramers' opacity $m=-7/2$. Simulations show that $\alpha$ increases when the magnetic Prandtl number approaches unity, and earlier work by \cite{2008ApJ...674..408B} calculated that the magnetic Prandtl number is expected to approach unity at a distance of $\sim 50$ Schwarzschild radii in a thin disc with physical parameters appropriate for an X-ray binary. Assuming a simple power law dependence $\alpha \propto \Pm^{n}$, then in the case of Coloumb viscosity $\alpha \propto T^{4n}_{c}/\rho^{n}$.   Using (\ref{radequi}) and (\ref{instkram})  and assuming electron scattering, one finds the instability criterion
\be
 n> 2/3 \  \  \m {(Instability, \ Coloumb\ viscosity, electron\ scattering)}. 
\label{instcrit}
\ee
This is in agreement with \cite{2011ApJ...727..106T}. Simulations suggest that $\alpha$ does indeed increase when $\Pm$  passes through unity.  The power law index  lies between $0.25\lesssim n \lesssim 0.9$, depending on the simulation (\citealt{2010A&A...516A..51L} and \citealt{2011ApJ...730...94S}) though there is no explicit temperature dependence of Pm in the simulations. This result suggests an avenue of exploration for the observed transition between spectral states in X-ray binaries. In section 6 we investigate the global behaviour of the instability using a simple time-dependent numerical simulation of a thin disc with a variable $\alpha$. 

Let us now calculate the instability criterion in the more general case of a power-law form for the opacity $\tau \propto \Sigma \rho T_{c}^{m}$, with $\alpha \propto \Pm^{n}$ and Coulomb viscosity.  Substituting $\alpha$ and $\tau$ into (\ref{radequi})
\be
T_{c}\propto \Sigma^{\frac{6-2n}{7-2m-9n}}, 
\ee
\be
T_{\m{s}}\propto \Sigma^{\frac{13+9n-2m+2mn}{4(7-2m-9n)}}.
\ee
So our disc instability criterion (\ref{unstable2}) becomes
\be
\frac{13+9n-2m+2mn}{4(7-2m-9n)}<0.
\label{inst}
\ee
Taking the opacity now to be given by Kramers' Law ($m=-7/2$), the instability criterion for $\alpha \propto \Pm^{n}$ is
\be
 n> 14/9 \  \  \m {(Instability, \ Coloumb\ viscosity, Kramers\ opacity)}.
\label{instcrit2}
\ee

Whilst it seems theoretically likely that the magnetic Prandtl number affects the value of $\alpha$, note that MHD simulations lack the resolution required to probe the high Reynolds and magnetic Reynolds numbers of real astrophysical accretion discs. In addition to the dependence of $\alpha$ on $\Pm$, simulations have also found a weak dependence on the Reynolds number of the plasma \citep{2010A&A...514L...5F}.   
At this stage, it is best to treat the detailed results of these simulations, such as the value of the power law dependence of $\alpha$ on $\Pm$, as a guideline.    

 \subsection{Radiative viscosity}

At the high central temperatures occurring in the inner accretion disc the effects of radiative viscosity become important. 
In a standard thin $\alpha$-disc with a bound-free Kramers opacity, the values of the central temperature and density are \citep{2002apa..book.....F} 
\be
T_{c}=1.4\times 10^{4} \alpha^{-1/5}\dot{m}_{16}^{3/10}M_{1}^{1/4}R_{10}^{-3/4}f^{6/5}K,
\label{FKRTc}
\ee
\be
\rho=3.1\times 10^{-8} \alpha^{-7/10}\dot{m}_{16}^{11/20}M_{1}^{5/8}R_{10}^{-15/8}f^{11/5} \m{gcm}^{-3},
\ee
\be
\qquad f=\left[1-\left(\frac{R_{*}}{R}\right)^{1/2}\right]^{1/4},
\ee
where $\dot{m}_{16}=\dot{m}/10^{16} \m{gs}^{-1}$, $M_{1}=M/M_{\odot}$ and $R_{*}$ is the inner edge of the disc where the stress is taken to be zero. Inserting these values (\ref{nuc/nur}) becomes 
\be
\frac{\nu_{C}}{\nu_{r}}=0.033 \alpha^{-2/5}\dot{m}_{16}^{-2/5}f^{-8/5}, 
\ee
It is interesting to note that this ratio depends only on the accretion rate and $\alpha$, so for accretion rates $\dot{m}>2.0\times 10^{12}/\alpha \,\m{gs}^{-1}$, the radiative viscosity will be larger than the Coulomb viscosity throughout the entire disc. Converting lengths into units of Schwarzschild radii and accretion rate into units of Eddington luminosity this becomes  
\be
\frac{\nu_{C}}{\nu_{r}}=0.012 \alpha^{-2/5} \left(\frac{\dot{mc^{2}}}{L_{\m{Edd}}}\right)^{-2/5}\left(\frac{M}{M_{\odot}}\right)^{-2/5}f^{-8/5}.
\ee
So the radiative viscosity will dominate the Coulomb viscosity for accretion rates $\dot{m}c^{2} \gtrsim 4 \times 10^{-5}L_{\m{Edd}}$, assuming $M=10M_{\odot}$ and $\alpha=0.05$. At high accretion rates approaching the Eddington luminosity we expect electron scattering to dominate Kramers' opacity in the inner part of a disc. Calculating the ratio of Coulomb and radiative viscosities using typical thin disc parameters \citep{2002apa..book.....F} with $\kappa=0.4 \m{cm}^{2}\m{g}^{-1}$
\be
\frac{\nu_{C}}{\nu_{r}}=0.089 \alpha^{-2/5}\left(\frac{\dot{mc^{2}}}{L_{\m{Edd}}}\right)^{1/10}\left(\frac{M}{M_{\odot}}\right)^{-2/5}\left(\frac{R}{r_{s}}\right)^{-3/4}f^{2/5}.
\ee
This shows that the radiative viscosity is larger than the Coulomb viscosity in the inner region of the disc (where the assumption of a dominant electron scattering opacity is valid) with only a weak dependence on the accretion rate. These calculations show that it is likely that the radiative viscosity is larger than the Coulomb viscosity in discs with high accretion rates. It is therefore worthwhile to calculate the instability criteria in the case where the radiative viscosity is considerably larger than the Coulomb viscosity, to see how this affects our results. If the radiative viscosity dominates the Coulomb viscosity the magnetic Prandtl number is given by (\ref{Pmradvisc}). Inserting values for the central temperature and density appropriate for a standard thin disc with a bound-free Kramers' opacity the magnetic Prandtl number is
\be
\Pm=5.2\times 10^{3} \alpha^{3/10}\left(\frac{\dot{mc^{2}}}{L_{\m{Edd}}}\right)^{21/20}\left(\frac{M}{M_{\odot}}\right)^{3/10}\left(\frac{R}{r_{s}}\right)^{-9/8}f^{21/5}.
\ee
The critical radius $R_{c}$ at which the magnetic Prandtl number is equal to unity is then
\be
\frac{R_{c}}{r_{s}}=2.0\times 10^{3} \alpha^{4/15}\left(\frac{\dot{mc^{2}}}{L_{\m{Edd}}}\right)^{14/15}\left(\frac{M}{M_{\odot}}\right)^{4/15}f^{56/15}.
\ee
For discs accreting at a high rate this radius will occur between $100r_{s}-1000r_{s}$, approximately an order of magnitude larger than the critical radius calculated in the case where the Coulomb viscosity is larger than the radiative viscosity. In the latter case the critical radius is given by 
\be
\frac{R_{c}}{r_{s}}=43 \alpha^{4/45}\left(\frac{\dot{mc^{2}}}{L_{\m{Edd}}}\right)^{26/45}\left(\frac{M}{M_{\odot}}\right)^{4/45}f^{104/45},
\ee
in agreement with the expression calculated by \cite{2008ApJ...674..408B}. We expect the instability to occur in discs in which $\Pm$ becomes greater than unity, associated with a change in $\alpha$. Using typical parameters ($R=10r_{s}$, $M=10M_{\odot}$, $\alpha=0.05$) to estimate the critical accretion rate, $\dot{m}_{c}$, above which $\Pm \geq 1$ at small disc radii  
\be
\dot{m}_{c}=4.2 \times 10^{-3} \frac{L_{\m{Edd}}}{c^{2}}.
\label{mcrit}
\ee
For X-ray binary systems with $\dot{m}>\dot{m}_{c}$ we expect the $\Pm$ instability to occur. We have shown that radiative viscosity is likely to be dominant over the Coulomb viscosity in discs accreting at a high Eddington fraction and the affect of this is to make the Prandtl number a sensitive function of temperature. This effectively increases the radius at which $\Pm=1$.  Let us now calculate the effect of radiative viscosity on the instability criteria in the case where the dominant opacity is electron scattering and $\alpha \propto \Pm^{n}$. Using equations \ref{unstable2}, \ref{tc} and \ref{radequi}
\be
T_{c}\propto \Sigma^{4(1-n)/(6-13n)}, \qquad T_{S}\propto \Sigma^{(10-3n)/(4(6-13n))},
\ee
$\m{(Instability, radiative\ viscosity, electron\ scattering)}$
\be
6/13<n<10/3.
\ee
The effect of a dominant radiative viscosity is to lower the instability threshold for $n$ compared to the case of Coulomb viscosity (\ref{instcrit}). Calculating the instability criteria in the case of a power law form for the opacity $\tau \propto \Sigma \rho T_{c}^{m}$
\be
T_{c}\propto \Sigma^{6(1-n)/(7-14n-2m+2mn)},
\ee
\be
T_{S}\propto \Sigma^{(13+n-2m+2mn)/(4(7-14n-2m+2mn))},
\ee
for which disc instability criterion is
\be
\frac{13+n-2m+2mn}{4(7-14n-2m+2mn)}<0.
\ee
In the case of a dominant Kramers\rq{} opacity, $m=-7/2$, this becomes\\ \\
$(\m{Instability, radiative\ viscosity, Kramers\ opacity})$
\be
2/3<n<10/3.
\label{instcrit4}
\ee
The effect of a dominant radiative viscosity is again to lower the instability condition for $n$ compared to the case of a dominant Coulomb viscosity (\ref{instcrit2}). 
In addition to the radiative viscosity there is also an analogous, radiative resistivity, caused by the deceleration of charged particles through Thompson scattering the radiation field. Unlike the radiative viscosity however, the radiative resistivity is far smaller than the Coulomb resistivity for plausible disc parameters (see Appendix). 

We now turn our attention to the behaviour of the instability in a simulated disc to understand how the instability manifests itself. 
					

\section{Numerical work}

In the computations discussed here we generally neglect the effect of radiation pressure. This is not because its effects are unimportant; indeed, questions of the induced instabilities go back to the early days of disc theory  (\citealt{1974ApJ...187L...1L} and \citealt{1976MNRAS.175..613S}).  But the precise contribution of radiation pressure to the stress (\ref{Trphi}) is still not well determined and the subject of ongoing research (\citealt{2009ApJ...691...16H}, \citealt{2009ApJ...704..781H} and \citealt{2011MNRAS.414.2186J}).   We shall assume that the opacity of the disc is the sum of the bound-free Kramers' opacity and electron scattering opacity, and thus that the electron scattering opacity dominates the inner sections of the disc.  When computing the value of the magnetic Prandtl number, the dynamic viscosity is taken as the sum of the Coulomb and radiative viscosities.

\subsection{Functional form for $\alpha$}

We adopt the following parameterization of $\alpha$ with $\Pm$

\be
\alpha(\Pm)=\alpha_{\m{min}}+(\alpha_{\m{max}}-\alpha_{\m{min}}) \left(\frac{\Pm^{u}}{\Pm^{u}+1}\right),
\label{Pmfunc}
\ee
where $\alpha_{\m{min}}$ and $\alpha_{\m{max}}$ are the minimum and maximum values of $\alpha$. This functional form tends smoothly to two distinct asymptotic values for $\alpha$, and at low-to-intermediate values of $\alpha$, $\Pm$ behaves approximately as $\alpha\simeq \alpha_{\m{min}}+(\alpha_{\m{max}}-\alpha_{\m{min}}) \Pm^{u}$.   Thus, the critical values of $u$ and $n$ should be comparable (\ref{instcrit}). Though very similar to a power law, the gradient of the adopted function is never quite as steep as the power law $\alpha \propto \Pm^{n}$ used to calculate the instability criteria, hence we expect the critical value of $u$ to be larger than the critical value of $n$.  The form of the expression (\ref{Pmfunc}) is not based on a deeper or more fundamental theory, and should be regarded as heuristic.

\subsection{Testing the instability criterion numerically}

\begin{figure*}
	\centering
		\subfloat[]{ \includegraphics[width=8.5cm, clip=true, trim=1.9cm 2cm 2cm 2cm]{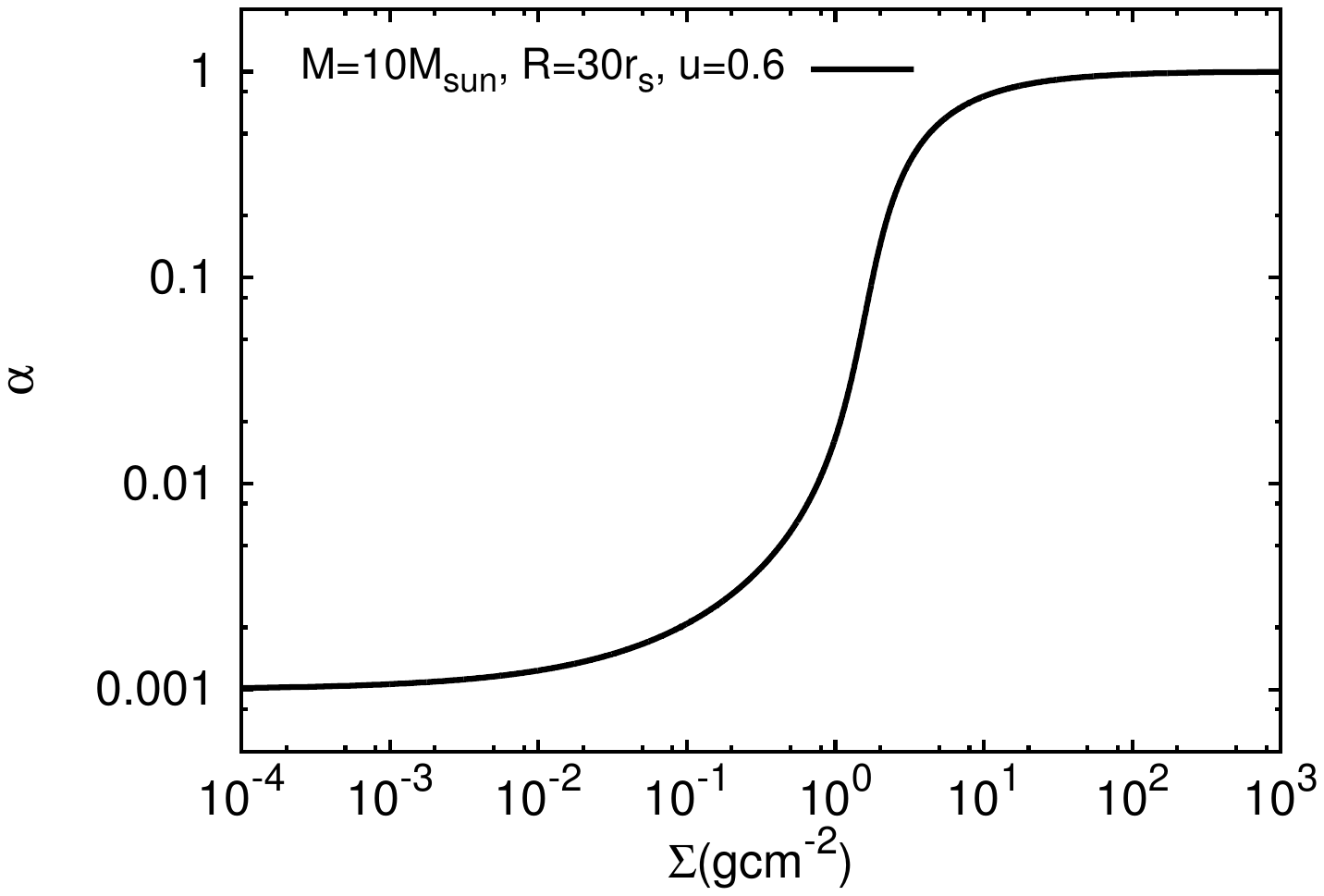} }
		\subfloat[]{ \includegraphics[width=8.5cm, clip=true, trim=1.9cm 2cm 2cm 2cm]{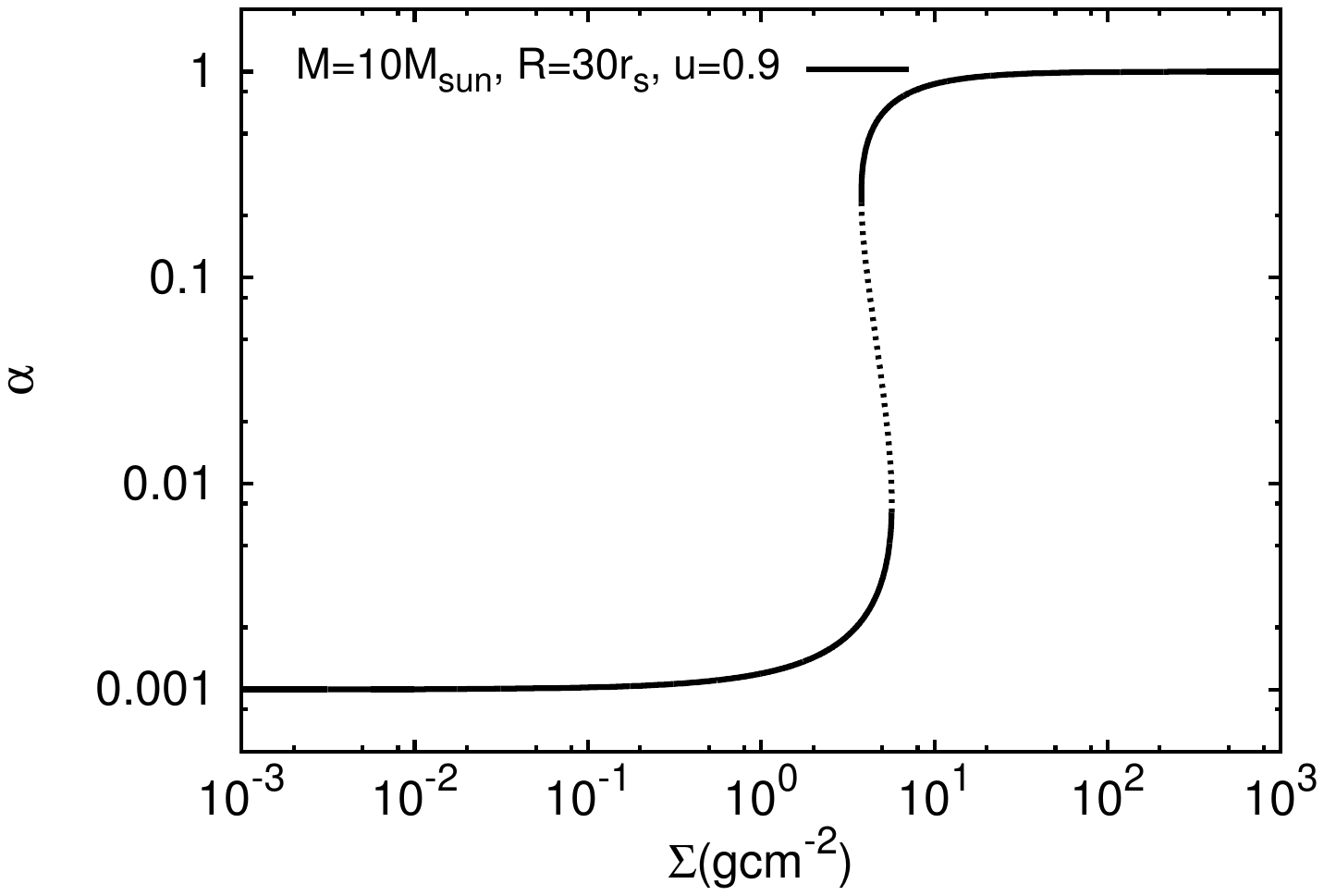} }	
		\\
		\subfloat[]{ \includegraphics[width=8.5cm, clip=true, trim=1.9cm 2cm 2cm 0cm]{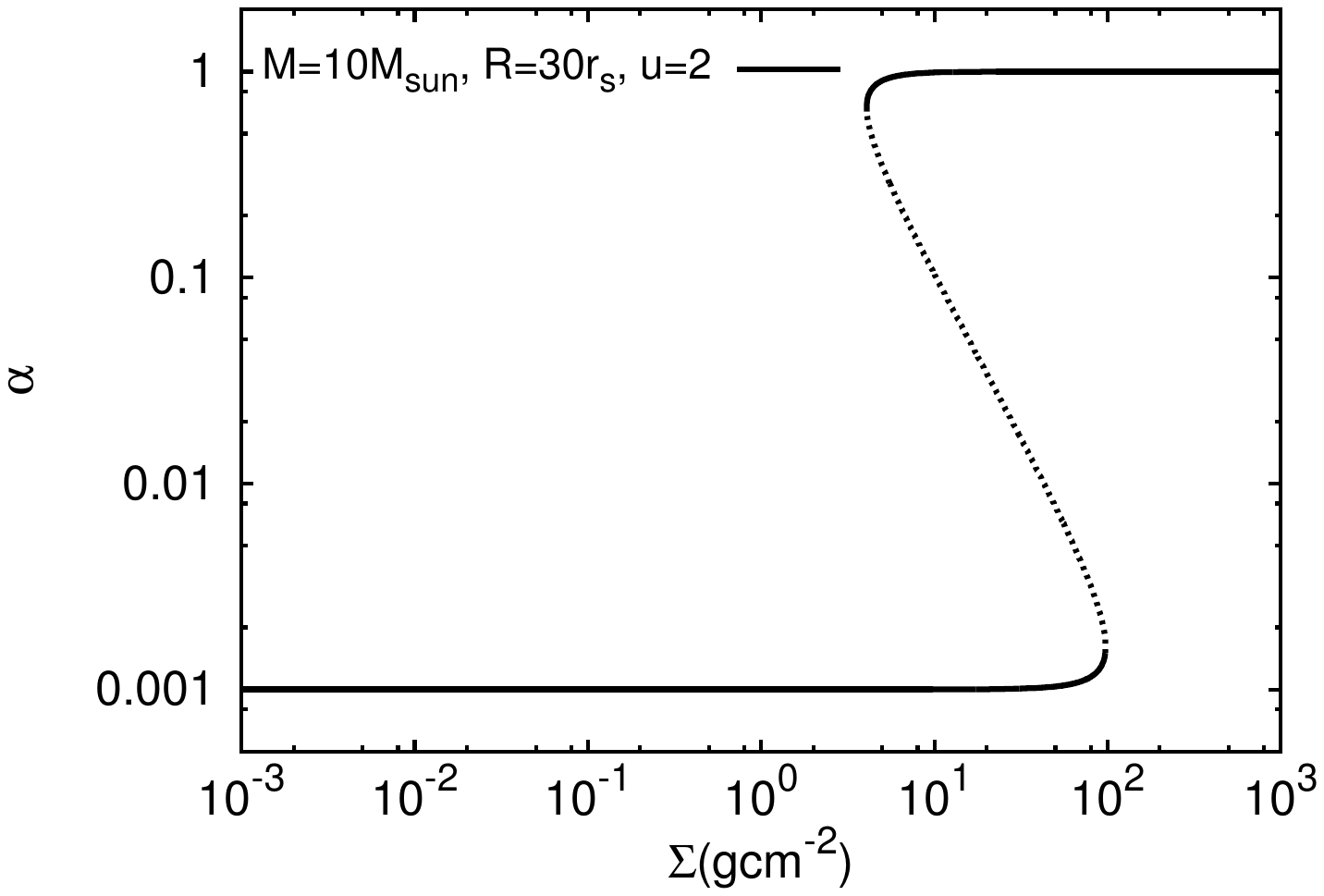} }
		\subfloat[]{ \includegraphics[width=8.5cm, clip=true, trim=1.9cm 2cm 2cm 2cm]{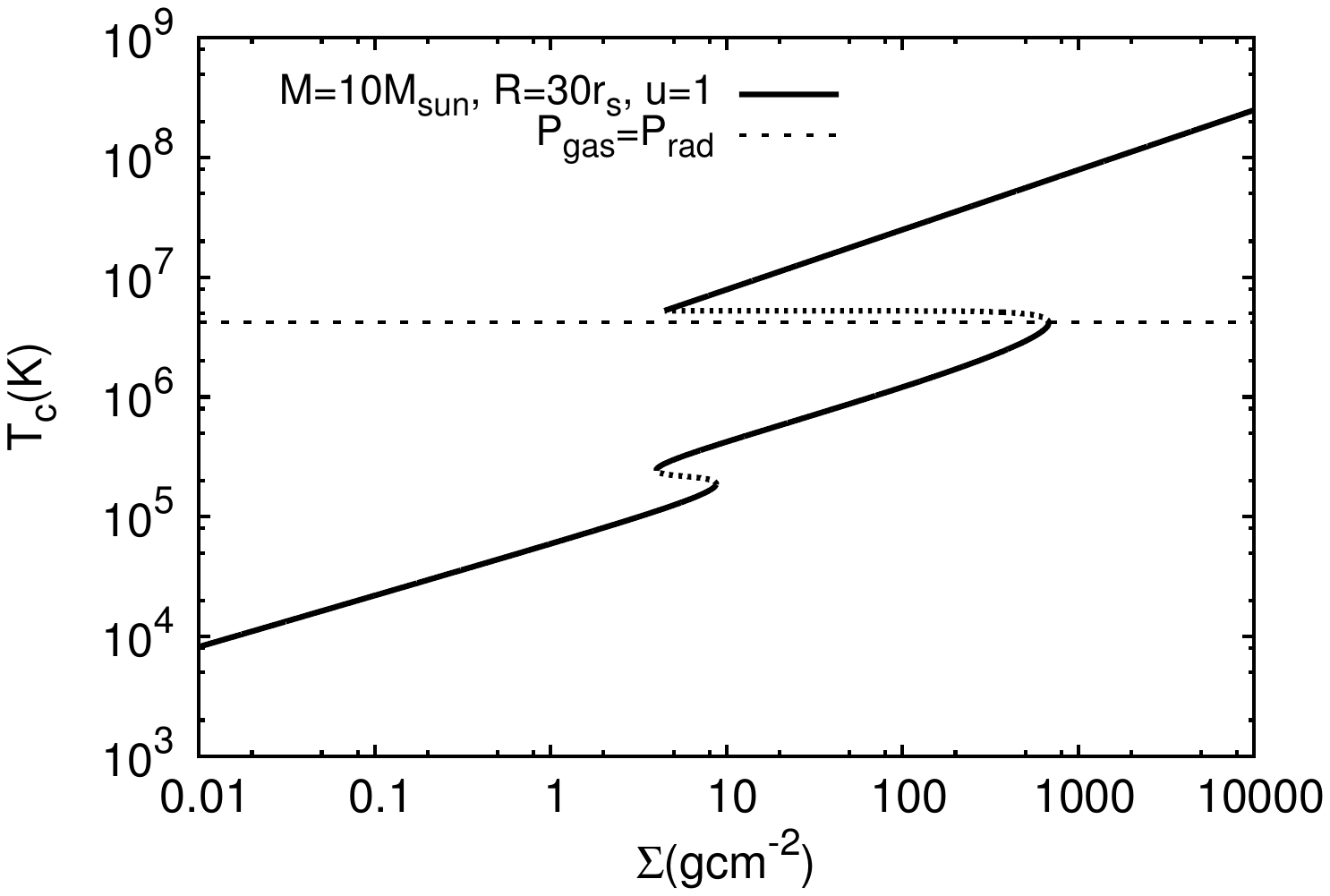} } 
					
	\caption{({\emph a}--{\emph c}): Solutions to (\ref{Disktemp}) with the prescription (\ref{Pmfunc}) for different values of $u$.   The central mass is $10M_{\odot}$; the location $r=30r_{s}$.   Radiation pressure is neglected.  If $u \gtrsim 0.8$, the solution behaves as an unstable limit cycle with two overlapping branches of stable solutions and a physically inaccessible region (dotted line). ({\emph d}): Central temperature $T_{c}$ as a function of $\Sigma$, {\em including both radiation and gas pressure.}  Additional radiation pressure dominated solutions are permitted; however, for these high $T_c$  solutions the thin disc assumption (H$\ll$R) breaks down. }
	\label{limcyc}
\label{nm}
\end{figure*}

The temperature of a thin disc in thermal equilibrium is given by the balance of local heating and cooling rates (\ref{eins})
\be
\frac{3}{2} HT_{R\phi}\Omega=\sigma T_{s}^{4}
\label{Disktemp}
\ee
where $T_{R\phi}$ is defined in (\ref{Trphi}) and $\alpha$ is given by (\ref{Pmfunc}). The solutions to (\ref{Disktemp}) are shown in Figures \ref{limcyc}a, b and c, for several values of $u$. For values of the index $u\gtrsim 0.8$, there are multiple-valued solutions for $\alpha$ (and $T_{c}$) for a given value of $\Sigma$. This value of $u$ is very similar to the criterion derived for the pure power law form for $\alpha$ in (\ref{instcrit4}) as expected. The unstable branch of solutions always satisfies our general instability criteria (\ref{geninst}). The figure suggests limit cycle behaviour for the disc instability, analogous to that induced by the changing opacity in DN systems. For the case $u\gtrsim 0.8$, there is a region in which three possible solutions exist with the lower branch of solutions extending to small values of $\Sigma$ and the upper branch extending to large $\Sigma$. This means that the value of $\alpha$ jumps discontinuously from the lower to upper branch of solutions as $\Sigma$ increases continuously from small to large values. If $u>1$, the solutions overlap over a significant range of values of $\Sigma$ so the value of $\alpha$ (and the associated accretion rate) remains on the upper branch even if $\Sigma$ then decreases below its original value. This system thus exhibits hysteresis.


The solutions to (\ref{Disktemp}) for $T_{c}$ as a function of $\Sigma$ including both gas and radiation pressure in the disc are plotted in Figure (\ref{limcyc}d). This solution exhibits a more complex structure than the case of a gas pressure disc. The additional branches of solutions arise from the rapid change in the sound speed when the pressure changes from the gaseous to radiative regime. The upper branch of radiative solutions  corresponds to the sound speed becoming comparable to the rotational velocity. The disc can no longer be considered geometrically thin under these circumstances, and in reality an optically thin ADAF-like solution (\citealt{1995ApJ...452..710N}) may be a better description of the flow. 




\section{Time-dependent disc simulation}

We have carried out a simple 1D time-dependent simulation of a thin accretion disc in which the $\alpha$ parameter depends on the magnetic Prandtl number. The disc is divided into logarithmically separated radial bins and the disc diffusion equation (\ref{Diff}) is solved using a finite difference scheme with adaptive timesteps. We account for the hysteresis of the solution by evaluating (\ref{Disktemp}) using the value of $\alpha$ from the previous timestep. Physically this accounts for the delay, of order the thermal timescale, over which changes occur to the disc temperature, and numerically this prevents an arbitrary choice of solution when multiple temperature equilibria exist for a given $\Sigma$. The outer boundary condition is taken to be the analytic solution to the thin disc equations and the inner boundary condition to be $\Sigma=0$ for $R<3r_{s}$. Many other plausible inner boundary conditions have been suggested, however, we have chosen our inner boundary condition for simplicity, since the density of the disc is expected to rapidly decrease within the innermost stable circular orbit for the black hole. We choose disc parameters $M=10M_{\odot}$, $\dot{M}=10^{-5}L_{Edd}/c^{2}$, $\alpha_{min}=0.001$ and $\alpha_{max}=1$ to produce figures (\ref{Disc}) and (\ref{Spec}). 

\subsection{Behaviour of the instability}

\begin{figure*}
	\centering

		 \includegraphics[width=12cm, clip=true, trim=0cm 0cm 0cm 0cm]{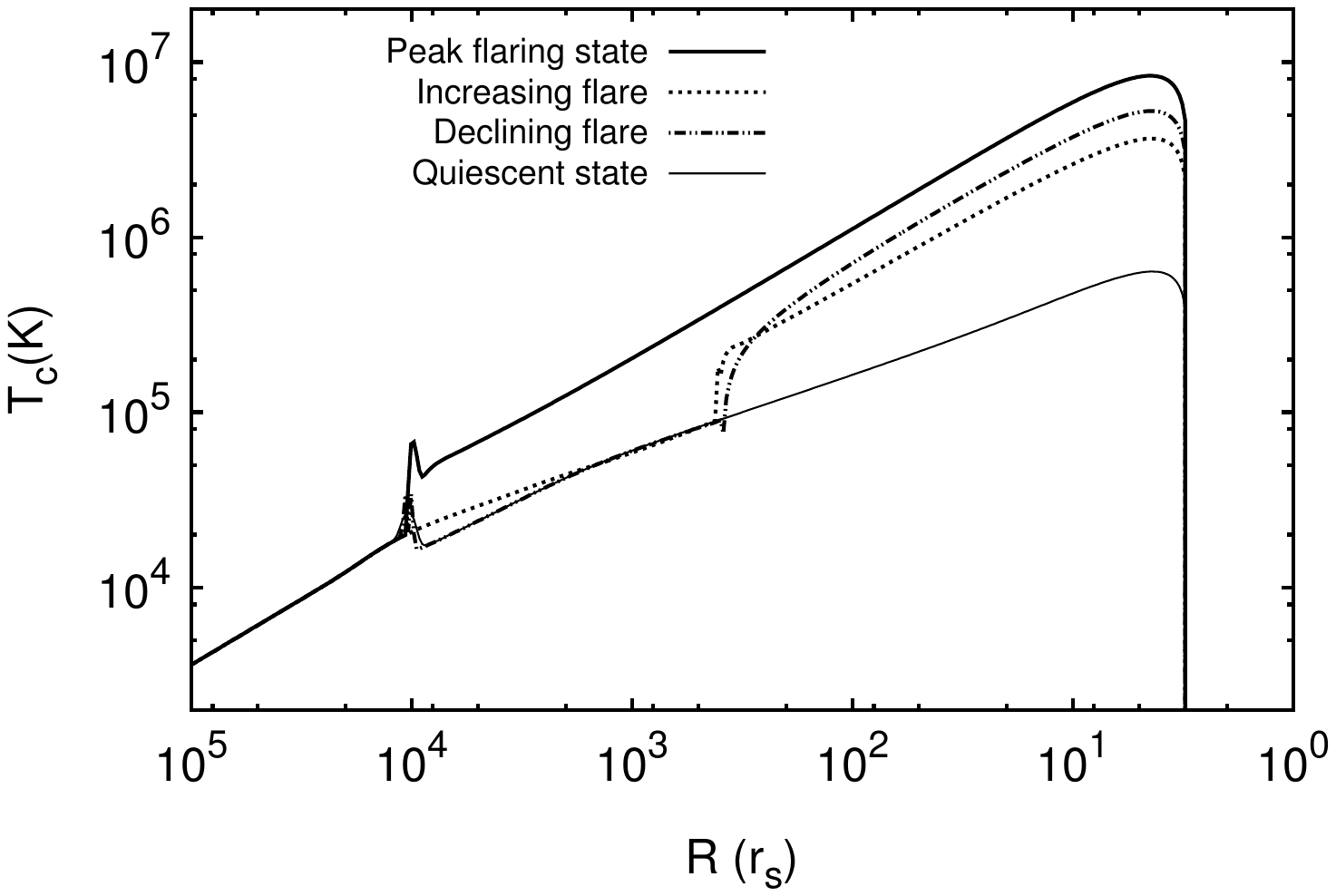} 
	\caption{The central temperature of the disc as a function of radius for the quiescent, flaring and intermediate states, with $u=2.0$, from our 1D simulation.}
	\label{Disc}
\end{figure*}

The instability exhibits remarkable behaviour in a global context.  When $\Pm$ approaches unity (for $u>0.8$), the value of $\alpha$ jumps discontinuously from small to large values and this causes the local accretion rate and temperature of the region to increase (recall $\dot{M} \propto \Sigma \alpha c_{s} H$). The instability is not localised to the initial triggering region however, since increasing the accretion rate in one zone must increase the rate in neighbouring zones. These are coupled via the diffusion equation (\ref{Diff}). The neighbouring zones may then become unstable themselves, should the increased accretion rate and heightened temperature be sufficient to increase the local $\Pm$ through unity. This causes the $\alpha$ parameter to jump to the upper branch of solutions (see Figure [\ref{limcyc}]). The increased accretion rate (and value of $\alpha$) will continue to propagate outwards from the initially unstable region until it reaches a radius at which the induced increase in accretion rate and surface density $\Sigma$ are insufficient to force $\alpha$ to move to the upper branch.  At this point, the increased accretion rate in the inner region of the disc becomes unsustainable. Due to a smaller value of $\alpha$, the accretion rate in the outer region of the disc is now below that of the unstable interior region. The discrepant accretion rates cause the mass in the inner region to be drained away.  The cycle repeats, however, once the mass in the depleted inner regions (now with $\alpha$ on the {\em lower} branch of solutions) has been replenished by accretion from the outer parts of the disc.   The Prandtl number $\Pm$ then becomes large enough to cause $\alpha$ to increase discontinuously in the inner disc, and the cycle repeats. 

This is a classic limit cycle.  A quiescent period, in which the depleted inner regions accrete mass, is followed by a hot state, in which the value of $\alpha$ locally jumps to the upper branch of solutions. This causes the accretion rate and temperature in the region to suddenly increase.  The instability then propagates from the hot inner region of the disc outwards (to $R\sim 10^{4}r_{s}$  in the simulations; this radius depends on the accretion rate) where it stalls.  A region of depleted mass then propagates back inwards, emptying the inner part of the disc. The stages of this process are shown in Figure \ref{Disc}. The detailed behaviour of the instability depends on the parameters in our chosen function (\ref{Pmfunc}). The instability propagates with a speed of order the turbulent diffusion timescale $R^2\Omega/W_{R\phi}$.   The relative fraction of time spent in the flaring and quiescent states is then $\sim \alpha_{min}/\alpha_{max}$. 

The parameter $u$ in (\ref{Pmfunc}) determines whether the disc is unstable, as well as the \lq\lq{}noisiness\rq\rq{} of the transition. Large values of $u$ result in a clean cyclic behaviour in which the disc instability propagates outwards in one wave, followed by a wave of depleted mass propagating back inwards. For unstable values of $u$, close to the critical value $u\approx0.8$, the edge of the unstable region propagates outwards and inwards many times before the inner region becomes depleted and enters the quiescent state. The noisiness is determined by the width of the region containing multiple values of $\alpha$. If this region is large (for large $u$) the upper branch of solutions for $\alpha$ exist down to lower values of $\Sigma$. This means that when $\alpha$ jumps to the upper branch of solutions it is more stable to a decrease in $\Sigma$.   

For values of $u$ such that multiple $\alpha$ solutions exist, we find that the gradient $\partial (\ln \alpha)/\partial (\ln \Sigma)$ of the middle branch of solutions always satisfies the criterion for instability (\ref{geninst}).   In our simulations the disc instability manifests itself as a limit cycle in $\alpha$, with the value of $\alpha$ jumping discontinuously from a lower branch of solutions to an upper branch. This is not quite the same physical behaviour as suggested by the initial instability criterion in (\ref{instkram}). This was a criterion for an unstable (but continuous) increase in the surface density of a region. In practice, regions which satisfy the criterion for unstable growth always have multiple solutions of $\alpha$ (and correspondingly $T_{c}$) forming a limit cycle. This is because $T_{s}$ should always increase with $\Sigma$ in the disc, and thus a region with $\partial T_{s}/\partial \Sigma<0$ must be multiple valued to ensure that $T_{s}$ always increases with $\Sigma$ for the physically accessible paths. The solutions formally satisfying the instability criterion are physically inaccessible for a disc in thermal equilibrium, so the instability in our simulations is due solely to the unstable limit cycle behaviour. 

The existence of the unstable behaviour requires that the equilibrium surface density of at least one part of the disc exceed the maximum value of the lower branch of solutions.  Thus, a disc with a low surface density at all radii will be stable, and may explain why not all X-ray binary systems are observed to have cyclic flaring behaviour. The accretion rate above which the instability exists, i.e. $\Pm>1$ in the disc, is difficult to calculate precisely since the magnetic Prandtl number is very sensitive to the temperature and density, $\Pm\propto T^{9}\rho^{-3}$. This is why our numerical simulation of a disc with an accretion rate $10^{-5}L_{\m{Edd}}$ is unstable, whilst the estimate (\ref{mcrit}) would suggest the disc should be stable (because the temperature we calculate numerically differs by a factor $\sim2$ from [\ref{FKRTc}]).   

\section{Comparison with observations}

\begin{figure*}
	\centering

		 \includegraphics[width=12cm, clip=true, trim=0cm 0cm 0cm 0cm]{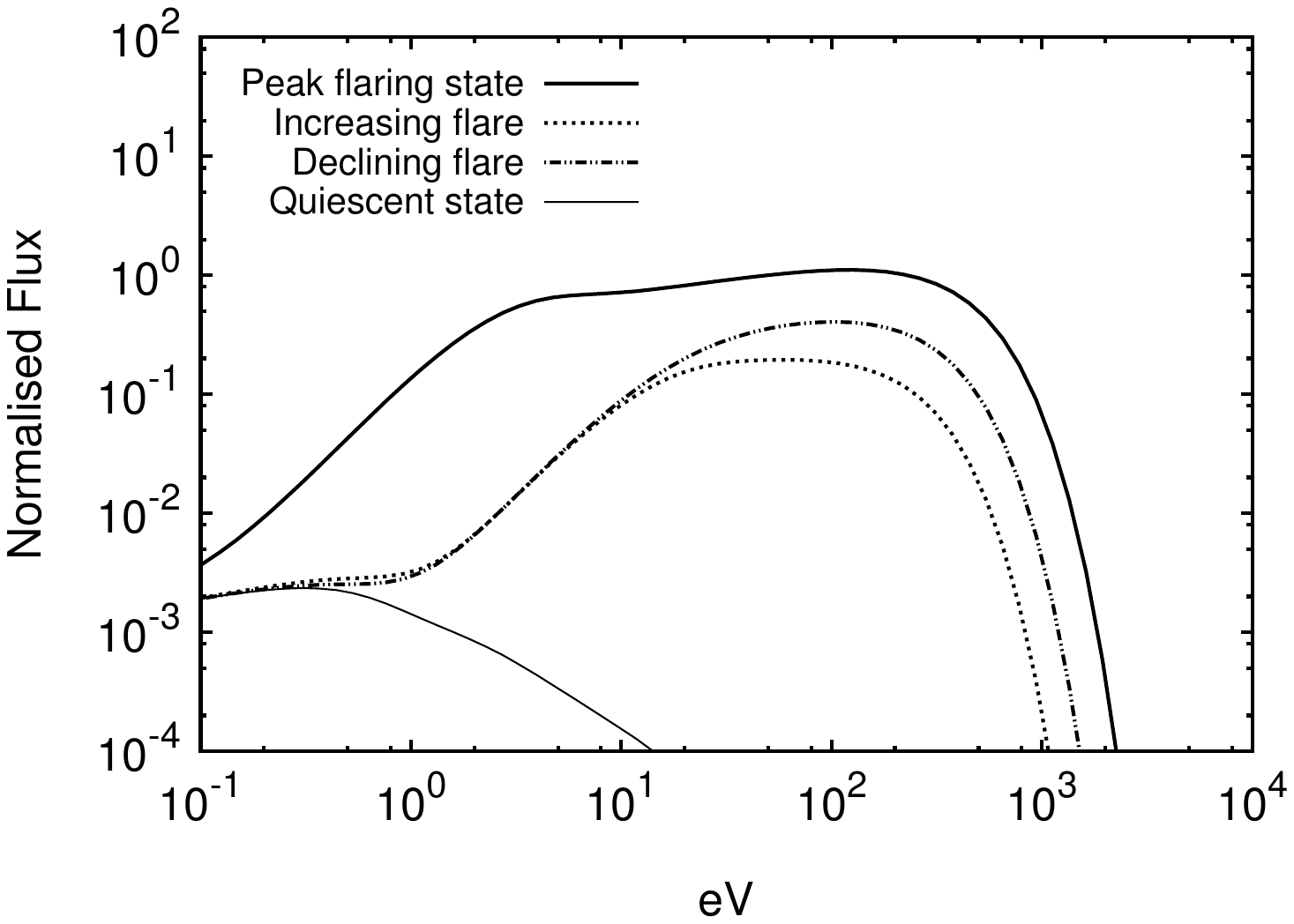} 
	\caption{The emitted spectrum in the quiescent, flaring and intermediate disc states (as shown in Figure \ref{Disc}).}
	\label{Spec}
\end{figure*}

The motivation for this work is the observed transitions of the spectral states of X-ray binaries. It has long been observed that X-ray binaries exhibit different spectral states which change cyclically over time. The states are broadly characterised by a high/soft state, a low/hard state, a very high/hard state and a quiescent state.  Most of the time, the disc is in quiesence ($\sim$years), periodically flaring over a period of weeks/months \citep{2006ARA&A..44...49R}. The physical mechanism behind these transitions is currently not well understood.   Switching from an efficiently radiating to inefficiently radiating solution has been viewed as promising path \citep{1995ApJ...452..710N}, though what prompts the transition itself remains to be elucidated.    

Current modelling holds that the outer disc is a standard thin disc, which extends to small radii in the high/soft state.   The hydrogen ionisation disc instability familiar from DN models is then invoked to produce the unstable flaring behaviour. The optically thick disc becomes unstable at large radii ($\sim 10^{10}$cm), with a heating front propagating both inwards and outwards (see \citealt{2001NewAR..45..449L} and references therein).   In this scenario, the heating front travels to the outer edge of the disc where it stalls, mass is then depleted from the inner disc causing it to become truncated. Irradiation of the disc is thought to maintain the disc in its high state resulting in the observed exponential decline of flares \citep{1998MNRAS.293L..42K}. Inside the truncation radius of the optically thick disc the low density material is thought to become hot and optically thin, producing hard X-rays and corresponding to the low/hard state. In addition, it is usually required that the value of $\alpha$ in the flaring state is larger by an order of magnitude than in the quiescent state, although there is no physical reason for this. These models are broadly consistent with observations \citep{2007A&ARv..15....1D}.

The behaviour of the instability outlined in this paper is similar to that of current X-ray binary models based on the DN hydrogen ionisation instability, but only because of the mathematical similarities of the two instabilities.  Unlike the hydrogen ionisation instability which occurs in the cool outer parts of a disc ($10^{4}<T<10^{5}$K), the instability presented here is caused by a variation of $\alpha$ with magnetic Prandtl number due to fundamental MHD (the difficulty of magnetic reconnection when the viscosity is large), and originates in the hot, inner disc. The flaring disc region will naturally have a larger $\alpha$ than the quiescent disc, since this is the ultimate cause of the instability. This provides a physical basis for the different values of $\alpha$ in the flaring and quiescent states which are normally assumed in hydrogen ionisation instability models of flaring X-ray binaries, in order to be compatible with observations.

In this paper we have shown that if $\alpha$ depends on the magnetic Prandtl number (as suggested by simulations), then an instability can naturally exist in the accretion disc which produces flaring and quiescent disc states.   If the regime is dominated by radiative viscosity and electron
scattering, then the instability is triggered if $\alpha \propto \Pm^{n}$ and $6/13<n<10/3$. Figure \ref{Spec} shows the spectrum of the disc in the different states corresponding to those in Figure \ref{Disc}. In the quiescent state the inner disc is depleted of mass and the disc spectrum is cool, since the outer parts of the disc dominate the observations. As the inner disc refills with mass, the observed spectrum becomes brighter and hotter due to the increase in temperature and accretion rate of the inner disc. Once the inner regions of the disc become unstable, the spectrum increases dramatically in X-ray flux and flares because of the increased accretion rate of the inner part of the disc. The spectrum then transitions back to the initial quiescent state once the mass in the inner part of the disc has been accreted onto the black hole. In this paper we are primarily concerned with the behaviour of instability in an optically thick, geometrically thin disc.  Thus, we have only modelled the soft, thermal component of the disc spectrum. Further work is required in order to model the source of hard X-rays (currently thought to be produced by a low density, hot, optically thin accretion flow) and to include the effects of disc irradiation in order to make detailed comparisons to observations.

This scenario may be able to account for the variation in the observed spectral transitions in X-ray binaries. The timescales of the quiescent and flaring states are proportional to the viscous timescales which depend on the values of $\alpha_{min}$ and $\alpha_{max}$. The noisiness of the flaring activity depends on the value of $u$ in our numerical simulations, with smaller unstable $u$ producing noisier disc state transitions. This flexibility is required to reproduce the different qualitative behaviour of the spectral transitions in different X-ray binaries and could help to constrain the value of physical parameters, such as $\alpha$, in accretion discs.  Note that $\alpha$ can be interpreted as the ratio of magnetic to thermal energy densities. In the high state $\alpha\approx \alpha_{\m{max}}$, the increased accretion rate and temperature are accompanied by a corresponding increase in the magnetic field strength in the disc. This large field strength may lead to a more powerful disc corona which would increase the amount of non-thermal particle acceleration. 

\section{Conclusion}

In this paper we have investigated an important consequence of a variable $\alpha$ parameter, depending on the physical conditions of a plasma. We have shown that an $\alpha$ parameter which increases with the magnetic Prandtl number can cause an accretion instability analogous to the instability in dwarf novae discs. We have calculated the general condition for this instability to occur in a standard thin disc.  For either a bound-free Kramers' opacity or electron scattering opacity the instability criterion is $\partial \ln \alpha/ \partial \ln \Sigma<-5/4$. When $\alpha$ depends on the magnetic Prandtl number as $\sim \Pm^{n}$, the inner region of the disc, dominated by electron scattering and radiative viscosity, is unstable if $6/13<n<10/3$.

We have investigated the global behaviour of this instability in X-ray binary systems via a 1D time-dependent simulation of a thin accretion disc.  When $\alpha$ satisfies the instability criterion, we find limit cycle behaviour. This induces cyclic flaring in the disc as the value of $\alpha$ cycles between small and large values.  In the high state, an unsustainable increase in accretion rate of the inner region of the disc drains mass for ($R \lesssim 10^{4}r_{s}$), truncating the disc. This is followed by a quiescent period in which this mass is replenished.  The instability seems able to reproduce at least some of the thermal behaviour of flares in X-ray binaries. In the quiescent state, the inner part of the disc is cool, whilst in the flaring state the accretion rate and temperature in the disc increase substantially producing X-ray emission. The quiescent timescale depends on the turbulent diffusion timescale of the disc.

The discovery of what seems to be a physically-motivated instability that bears on observations of flaring in X-ray binary accretion discs is significant, since this is a long-standing problem. In the near future we will investigate whether this instability operates in three-dimensional MHD simulations of X-ray binary discs.

\section{Acknowledgements}

WJP acknowledges support from the University of Oxford. SAB acknowledges support from the Royal Society in the form of a Wolfson Research Merit Award. We would like to thank the anonymous referee for an exceptionally helpful report. 

\appendix
\section{Radiative resistivity}

The disc plasma is composed from electrons, ions, protons and photons. The ions and photons are coupled by bound-free scattering whilst the protons are coupled to the ions via Coulomb collisions. The resistivity of the plasma will be determined by the shortest electron deflection time (the electron mobility being significantly larger than that of the protons or ions). We are therefore interested in calculating the electron deflection times due to Coulomb scattering from ions and Thomson scattering photons. Electrons moving relative to the photon gas (with energy density $U_{\m{rad}}$) will experience a drag force due to inverse-Compton scattering the black body photons. The power emitted is given by \citep{2011hea..book.....L}
\be
P_{\ \m{IC}}=\frac{4}{3}\sigma_{T}c\beta^{2}\gamma^{2}U_{\m{rad}}
\ee
where $\sigma_{T}$ is the Thomson scattering cross section, $\beta$ is the relative velocity of the electron and photon gas divided by the speed of light and $\gamma$ is the Lorentz factor of the particle. The characteristic time for the electrons to decelerate by this process will be given by the ratio of the kinetic energy of the electrons to the energy radiated. The relative speed of the electrons and photon gas will be of order the thermal sound speed
\be
 t_{d \ \m{rad}}=\frac{m(\gamma-1)c^{2}}{P_{\ \m{IC}}}=\frac{mc_{s}^{2}}{2P_{\ \m{IC}}}=\frac{3m(c^{2}-c_{s}^{2})}{8\sigma_{T}U_{\m{rad}}}
\ee
where we have assumed that $\beta \ll 1$, the energy density of a blackbody photon gas is given by $U_{\m{rad}}=aT_{c}^{4}$ and the average rms particle velocity of a monatomic ideal gas is $mc_{s}^{2}=3kT_{c}$. The deflection time is then
\be
t_{d \ \m{rad}}=2.0\times10^{21} \left[\frac{1-3kT_{c}/(m_{e}c^{2})}{T_{c}^{4}}\right] \m{s}
\ee
The characteristic deflection time from Coulomb collisions of electrons with ions is given by (see \citealt{1962pfig.book.....S}, P138-139 and \citealt{2008ApJ...674..408B})
\be
t_{d \ \m{C}}=\frac{m_{e}^{1/2}(3kT_{c})^{3/2}F(x) \ln \Lambda}{8\pi n_{i}Z_{i}^{2}e^{4}}
\ee
\be
F(x)=\left(1-\frac{1}{2x^{2}}\right)\frac{2}{\sqrt{\pi}}\int_{0}^{x}e^{-s^{2}}\m{d}s+\frac{e^{-x}}{\sqrt{\pi}}, \ \ \ \ x=\sqrt{\frac{m_{i}v_{i}^{2}}{2kT_{c}}}
\ee
We take the value of the Coulomb logarithm to be $\ln \Lambda=\sqrt{40}$ and if we assume $x=\sqrt{3/2}$ for thermalised ions then $F(x)\ln \Lambda=4.9$. For a $90\%$ hydrogen, $10\%$ helium composition (by number) the weighted ion charge is $Z_{i}^{2}=1.3$ and the ion number density $n_{i}=0.91 \rho/m_{p}$. We estimate the time between Coulomb collisions of electrons with ions to be
\be
t_{d \ \m{C}}\approx7.2\times 10^{5}\frac{T_{c}^{3/2}}{n_{i}} 
\ee
The ratio of the two deflection times is then
\be
\frac{t_{d \ \m{rad}}}{t_{d \ \m{C}}}=2.8\times 10^{15} n_{i}T_{c}^{-11/2} 
\ee
Inserting an estimate of the disc temperature and density (\ref{FKRTc}) we find
\be
\frac{t_{d \ \m{rad}}}{t_{d \ \m{C}}}=7.4 \times 10^{11} \alpha^{2/5}\dot{m_{16}}^{-11/10}M_{1}^{-3/4}R_{10}^{9/4}f^{-22/5} 
\ee
Rewriting the accretion rate in terms of the Eddington luminosity and radius in Schwarzschild radii this ratio becomes
\be
\frac{t_{d \ \m{rad}}}{t_{d \ \m{C}}}=2.7 \alpha^{11/10}\left(\frac{\dot{mc^{2}}}{L_{\m{Edd}}}\right)^{-11/10}M_{1}^{13/5}\left(\frac{R}{r_{s}}\right)^{9/4}f^{-22/5} 
\ee
So the radiative resistivity will only become important in the innermost region of discs with super-Eddington accretion rates. For typical parameters associated with transient black hole binary systems the Coulomb resistivity will be significantly larger than the radiative resistivity (i.e. $t_{d \ \m{rad}} /t_{d \ \m{C}} \gg1$).

\bibliographystyle{mn2e}
\bibliography{PmInstability}

\begin{thebibliography}{}

\bibitem[\protect\citeauthoryear{{Balbus} \& {Hawley}}{{Balbus} \&
  {Hawley}}{1998}]{1998RvMP...70....1B}
{Balbus} S.~A.,  {Hawley} J.~F.,  1998, Reviews of Modern Physics, 70, 1

\bibitem[\protect\citeauthoryear{{Balbus} \& {Henri}}{{Balbus} \&
  {Henri}}{2008}]{2008ApJ...674..408B}
{Balbus} S.~A.,  {Henri} P.,  2008, \apj, 674, 408

\bibitem[\protect\citeauthoryear{{Balbus} \& {Papaloizou}}{{Balbus} \&
  {Papaloizou}}{1999}]{1999ApJ...521..650B}
{Balbus} S.~A.,  {Papaloizou} J.~C.~B.,  1999, \apj, 521, 650

\bibitem[\protect\citeauthoryear{{Done}, {Gierli{\'n}ski} \& {Kubota}}{{Done}
  et~al.}{2007}]{2007A&ARv..15....1D}
{Done} C.,  {Gierli{\'n}ski} M.,    {Kubota} A.,  2007, \aapr, 15, 1

\bibitem[\protect\citeauthoryear{{Faulkner}, {Lin} \& {Papaloizou}}{{Faulkner}
  et~al.}{1983}]{1983MNRAS.205..359F}
{Faulkner} J.,  {Lin} D.~N.~C.,    {Papaloizou} J.,  1983, \mnras, 205, 359

\bibitem[\protect\citeauthoryear{{Frank}, {King} \& {Raine}}{{Frank}
  et~al.}{2002}]{2002apa..book.....F}
{Frank} J.,  {King} A.,    {Raine} D.~J.,  2002, {Accretion Power in
  Astrophysics: Third Edition}

\bibitem[\protect\citeauthoryear{{Fromang}}{{Fromang}}{2010}]{2010A&A...514L...5F}
{Fromang} S.,  2010, \aap, 514, L5

\bibitem[\protect\citeauthoryear{{Fromang}, {Papaloizou}, {Lesur} \&
  {Heinemann}}{{Fromang} et~al.}{2007}]{2007A&A...476.1123F}
{Fromang} S.,  {Papaloizou} J.,  {Lesur} G.,    {Heinemann} T.,  2007, \aap,
  476, 1123

\bibitem[\protect\citeauthoryear{{Hirose}, {Blaes} \& {Krolik}}{{Hirose}
  et~al.}{2009}]{2009ApJ...704..781H}
{Hirose} S.,  {Blaes} O.,    {Krolik} J.~H.,  2009, \apj, 704, 781

\bibitem[\protect\citeauthoryear{{Hirose}, {Krolik} \& {Blaes}}{{Hirose}
  et~al.}{2009}]{2009ApJ...691...16H}
{Hirose} S.,  {Krolik} J.~H.,    {Blaes} O.,  2009, \apj, 691, 16

\bibitem[\protect\citeauthoryear{{Janiuk} \& {Czerny}}{{Janiuk} \&
  {Czerny}}{2011}]{2011MNRAS.414.2186J}
{Janiuk} A.,  {Czerny} B.,  2011, \mnras, 414, 2186

\bibitem[\protect\citeauthoryear{{King} \& {Ritter}}{{King} \&
  {Ritter}}{1998}]{1998MNRAS.293L..42K}
{King} A.~R.,  {Ritter} H.,  1998, \mnras, 293, L42

\bibitem[\protect\citeauthoryear{{Lasota}}{{Lasota}}{2001}]{2001NewAR..45..449L}
{Lasota} J.-P.,  2001, \nar, 45, 449

\bibitem[\protect\citeauthoryear{{Lesur} \& {Longaretti}}{{Lesur} \&
  {Longaretti}}{2007}]{2007MNRAS.378.1471L}
{Lesur} G.,  {Longaretti} P.-Y.,  2007, \mnras, 378, 1471

\bibitem[\protect\citeauthoryear{{Lightman} \& {Eardley}}{{Lightman} \&
  {Eardley}}{1974}]{1974ApJ...187L...1L}
{Lightman} A.~P.,  {Eardley} D.~M.,  1974, \apjl, 187, L1

\bibitem[\protect\citeauthoryear{{Loeb} \& {Laor}}{{Loeb} \&
  {Laor}}{1992}]{1992ApJ...384..115L}
{Loeb} A.,  {Laor} A.,  1992, \apj, 384, 115

\bibitem[\protect\citeauthoryear{{Longair}}{{Longair}}{2011}]{2011hea..book.....L}
{Longair} M.~S.,  2011, {High Energy Astrophysics}

\bibitem[\protect\citeauthoryear{{Longaretti} \& {Lesur}}{{Longaretti} \&
  {Lesur}}{2010}]{2010A&A...516A..51L}
{Longaretti} P.-Y.,  {Lesur} G.,  2010, \aap, 516, A51

\bibitem[\protect\citeauthoryear{{Meyer}, {Liu} \& {Meyer-Hofmeister}}{{Meyer}
  et~al.}{2000}]{2000A&A...361..175M}
{Meyer} F.,  {Liu} B.~F.,    {Meyer-Hofmeister} E.,  2000, \aap, 361, 175

\bibitem[\protect\citeauthoryear{{Mihalas} \& {Mihalas}}{{Mihalas} \&
  {Mihalas}}{1984}]{1984oup..book.....M}
{Mihalas} D.,  {Mihalas} B.~W.,  1984, {Foundations of radiation hydrodynamics}

\bibitem[\protect\citeauthoryear{{Narayan} \& {Yi}}{{Narayan} \&
  {Yi}}{1995}]{1995ApJ...452..710N}
{Narayan} R.,  {Yi} I.,  1995, \apj, 452, 710

\bibitem[\protect\citeauthoryear{{Pringle}}{{Pringle}}{1981}]{1981ARA&A..19..137P}
{Pringle} J.~E.,  1981, \araa, 19, 137

\bibitem[\protect\citeauthoryear{{Remillard} \& {McClintock}}{{Remillard} \&
  {McClintock}}{2006}]{2006ARA&A..44...49R}
{Remillard} R.~A.,  {McClintock} J.~E.,  2006, \araa, 44, 49

\bibitem[\protect\citeauthoryear{{Schekochihin}, {Cowley}, {Taylor}, {Maron} \&
  {McWilliams}}{{Schekochihin} et~al.}{2004}]{2004ApJ...612..276S}
{Schekochihin} A.~A.,  {Cowley} S.~C.,  {Taylor} S.~F.,  {Maron} J.~L.,
  {McWilliams} J.~C.,  2004, \apj, 612, 276

\bibitem[\protect\citeauthoryear{{Shakura} \& {Sunyaev}}{{Shakura} \&
  {Sunyaev}}{1973}]{1973A&A....24..337S}
{Shakura} N.~I.,  {Sunyaev} R.~A.,  1973, \aap, 24, 337

\bibitem[\protect\citeauthoryear{{Shakura} \& {Sunyaev}}{{Shakura} \&
  {Sunyaev}}{1976}]{1976MNRAS.175..613S}
{Shakura} N.~I.,  {Sunyaev} R.~A.,  1976, \mnras, 175, 613

\bibitem[\protect\citeauthoryear{{Simon} \& {Hawley}}{{Simon} \&
  {Hawley}}{2009}]{2009ApJ...707..833S}
{Simon} J.~B.,  {Hawley} J.~F.,  2009, \apj, 707, 833

\bibitem[\protect\citeauthoryear{{Simon}, {Hawley} \& {Beckwith}}{{Simon}
  et~al.}{2011}]{2011ApJ...730...94S}
{Simon} J.~B.,  {Hawley} J.~F.,    {Beckwith} K.,  2011, \apj, 730, 94

\bibitem[\protect\citeauthoryear{{Spitzer}}{{Spitzer}}{1962}]{1962pfig.book.....S}
{Spitzer} L.,  1962, {Physics of Fully Ionized Gases}

\bibitem[\protect\citeauthoryear{{Takahashi} \& {Masada}}{{Takahashi} \&
  {Masada}}{2011}]{2011ApJ...727..106T}
{Takahashi} H.~R.,  {Masada} Y.,  2011, \apj, 727, 106

\end{thebibliography}
\bibdata{PmInstability}

\label{lastpage}

\end{document}